# Astro2020 Science White Paper

## Ground-Based Radial Velocity as Critical Support for Future NASA Earth-Finding Missions

**Thematic Area:** Planetary Systems

**Principal Author:**
Name: Courtney D. Dressing
Institution: University of California, Berkeley
Email: dressing@berkeley.edu
Phone: 703-795-8799

**Co-authors:**
Christopher C. Stark, STScI, cstark@stsci.edu
Peter Plavchan, George Mason University, pplavcha@gmu.edu
Eric Lopez, NASA GSFC, eric.d.lopez@nasa.gov

**Co-signers:**
Aki Roberge, NASA/GSFC, Aki.Roberge@nasa.gov
Johanna Teske, Carnegie Observatories, jteske@carnegiescience.edu
John M. O'Meara, W. M. Keck Observatory, jomeara@keck.hawaii.edu
Leslie Rogers, University of Chicago, larogers@uchicago.edu
Eric Nielsen, KIPAC, Stanford University, enielsen@stanford.edu
Tyler D. Robinson, Northern Arizona University, tyler.robinson@nau.edu
Angelle Tanner, Mississippi State University, at876@msstate.edu
Jonathan Crass, University of Notre Dame, j.crass@nd.edu
Chuanfei Dong, Princeton University, dcfy@princeton.edu
Evgenya Shkolnik, Arizona State University, shkolnik@asu.edu
Henry Ngo, National Research Council Canada, Henry.Ngo@nrc-cnrc.gc.ca
William D. Cochran, University of Texas at Austin, wdc@astro.as.utexas.edu
Steve Bryson, NASA/ARC, steve.bryson@nasa.gov
Jason Wright, Penn State, astrowright@gmail.com
Arif Solmaz, Çağ University, arifscag@gmail.com
Eric B. Ford, Penn State, ebf11@psu.edu
B. Scott Gaudi, The Ohio State University, gaudi.1@osu.edu
Neil Zimmerman, NASA/GSFC, neil.t.zimmerman@nasa.gov
Lars A. Buchhave, DTU Space, National Space Institute, Technical University of Denmark, buchhave@space.dtu.dk
Rachel Akeson, Caltech-IPAC, rla@ipac.caltech.edu
David R. Ciardi, Caltech/IPAC-NExScI, ciardi@ipac.caltech.edu
David W. Latham, Harvard-Smithsonian Center for Astrophysics, dlatham@cfa.harvard.edu
Christopher Watson, Queen's University Belfast, c.a.watson@qub.ac.uk
Andrew Collier Cameron, University of St. Andrews, acc4@st-andrews.ac.uk

**Abstract**: Future space-based direct imaging missions are poised to search for biosignatures in the atmospheres of potentially habitable planets orbiting nearby AFGKM stars. Although these missions could conduct a survey of high-priority target stars to detect candidate Earth-like planets, conducting a precursor radial velocity (RV) survey will benefit future direct imaging missions in four ways. First, an RV survey capable of detecting signals as small as 8 cm/s over timescales of a few years could discover potentially habitable Earth-mass planets orbiting dozens of nearby GKM stars accessible to space-based direct imaging. Second, RVs will improve scheduling efficiency by reducing the required number of revisits for orbit determination, and revealing when a planet of interest is most observable. Third, RV observations will reveal the masses of gas and ice giants that could be mistaken for Earth-mass planets, thereby reducing the time spent identifying false positives. Fourth, mass measurements from RVs will provide the surface gravities necessary for interpreting atmospheric spectra and potential biosignatures.

**Introduction**
By discovering thousands of planets, *Kepler* revealed that planetary systems are common and that smaller planets are even more prevalent than larger planets (e.g., Howard et al. 2012, Fressin et al. 2013, Burke et al. 2015; within orbital periods of up to 200 days). Roughly 25% of M dwarfs and 2-34% of Sun-like stars harbor small planets within their habitable zones (HZ; Winn & Fabrycky 2015 & references therein). The potentially habitable planets discovered by *Kepler* are too distant for detailed study, but the *Kepler* occurrence rates indicate that dozens of potentially habitable planets should orbit the nearest stars. Due to geometry, most of these nearby planets do not transit and therefore will not be detected by transit surveys or be accessible via transmission and secondary eclipse spectroscopy. Accordingly, we must rely on other methods to find and study these potentially Earth-like planets. The NAS Exoplanet Science Strategy report recommended, *"NASA should lead a large strategic direct imaging mission capable of measuring the reflected-light spectra of temperate terrestrial planets orbiting Sun-like stars."* We endorse the findings and recommendations published in the National Academy reports on Exoplanet Science Strategy and Astrobiology Strategy for the Search for Life in the Universe. This white paper extends and complements the material presented therein.

Specifically, we discuss four reasons why **conducting a precursor radial velocity (RV) survey will benefit future direct imaging missions.** First, a precursor survey will enable *prioritization of target stars* by finding potentially habitable planets orbiting nearby GKM stars and measuring their masses. Second, RVs will *improve scheduling efficiency* by reducing the required number of revisits for orbit determination, and revealing when a planet of interest is at maximum elongation and thus favorable instrumental imaging contrast. Third, by determining the masses of gas and ice giants in the HZs of AFGKM stars, an RV survey will *reduce the time spent identifying false positives*. Fourth, planetary surface gravities are essential for *interpreting atmospheric spectra and potential biosignatures*.

**Targets Accessible to a Future Direct Imaging Survey**
For direct imaging studies, the set of stars that can be searched for planets is determined by the performance of the starlight suppression system (see Astro2020 White Paper (WP) by C. Stark for details). The three key parameters that define the possible target list are the inner working angle (IWA; the minimum angular separation at a planet can be detected), the outer working angle (OWA: the maximum angular separation), and the contrast (the minimum planet:star brightness ratio for detection). Improving the contrast will reveal planets that are fainter relative to their host stars and decreasing the IWA will yield planets at smaller angular separations and reduce the number of observations required to search the HZ of a given star (if orbits are known a priori). When using a coronagraph, the IWA is coupled to the diameter of the telescope such that larger telescopes are able to resolve planets at smaller angular separations. The OWA is determined by the size of the deformable mirror. The IWA and contrast depend on the specific choice of coronagraph, but for a given coronagraph the IWA will scale linearly with wavelength and inversely with telescope aperture (i.e., IWA $\propto \lambda/D$). Accordingly, planets that are in close proximity to their target stars may be accessible at bluer wavelengths yet inside the IWA (and therefore inaccessible) at redder wavelengths.

For starshade-based systems, the coupling between telescope aperture and IWA is weakened. Instead, the IWA is set by the diameter of the starshade and the distance between the starshade and telescope down to a practical limit of roughly $\lambda/D$. Planets observed using a starshade will therefore be accessible across the starshade's nominal bandpass range at the cost of spending the fuel required to move the starshade to the required position.

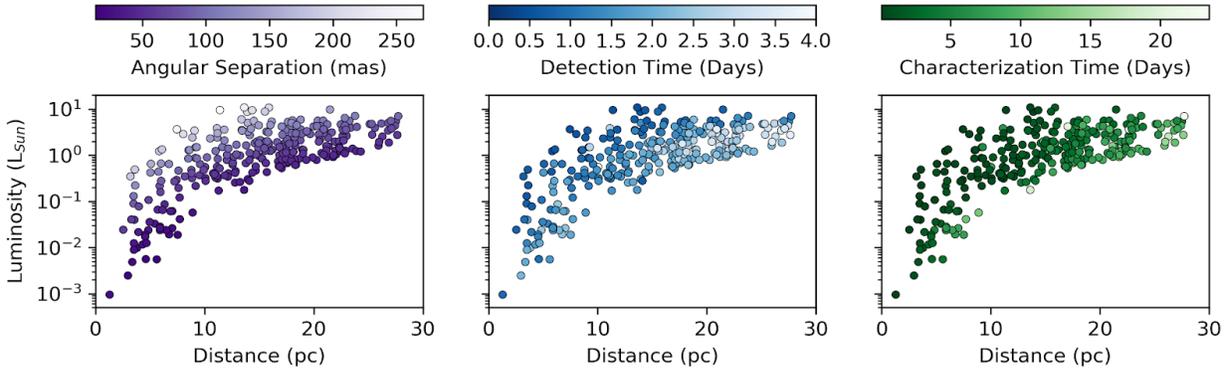

*Figure 1: Luminosity versus distance for stars with HZs accessible to a fiducial direct imaging mission. The colors indicate the angular separation of a planet receiving the same insolation flux as Earth (left), the time required for initial detection (middle) and the time required to characterize the planet by obtaining a spectrum covering all accessible wavelengths (right). Figure generated using simulations completed by Chris Stark (STScI).*

For the remainder of this WP, we adopt a 15-m NUV, optical, and NIR space telescope with a coronagraph based on the ECLIPS instrument developed for architecture A of the LUVOIR mission concept study. We selected this configuration because it has the largest potential target list of the direct imaging facilities currently under study, but our arguments can be applied to any future imaging mission. In *Figure 1*, we display the stars for which these simulations predict we could detect HZ planets. The target stars are bright (V = 2.68 - 11.12) and nearby (1.3 - 27.7 pc) with spectral types ranging from A2 to M5.5. A smaller observatory would be able to search for planets around the closest subset of these targets while a more powerful observatory would also survey more distant stars. The time required to detect each candidate Earth-like planets varies from 0.5 to 4 days (middle panel of *Figure 1*), although these observations will likely be split into multiple visits.

**Designing a Precursor RV Survey**

Although some of the stars in *Figure 1* may be too hot, quickly rotating, or active for precise RV observations with current facilities, their brightnesses are within the currently accessible range (see Fischer et al. 2016 for a review of RV capabilities). As noted by D. Ciardi (Astro2020 WP), the large apertures of ELTs will enable extremely short integration times and the introduction of more sophisticated strategies to measure the masses of planets orbiting active stars. Accordingly, even the more active stars may be amenable for future mass measurement.

Assuming that each of the stars in *Figure 1* is orbited by an Earth-mass planet receiving the same insolation flux as Earth, the anticipated RV signal for an edge-on orbital configuration would range from 1.4 m/s for the least massive star to 3.5 cm/s for the most massive star. In *Figure 2*, we display the simulated RV signal for each star as a function of stellar mass, planet orbital period, and V-band magnitude. Earth-like planets in the HZs of M and late K dwarfs will induce RV signals of 20 - 140 cm/s over orbital periods of 5 - 100 days, which should render them detectable with current and future facilities. For early K and G stars, the anticipated signals are smaller (5 - 20 cm/s) and occur over longer timescales (100 days - 2 years). As noted in the parallel Astro2020 WPs focused on radial velocity (Ciardi, etc.), the next generation of RV facilities including extremely precise spectrographs for use with the ELTs or space-based mission concepts like EarthFinder are being designed to find these planets. While achieving such a high effective precision will require advances in statistical analysis of high-resolution spectroscopy to discriminate planetary signals from intrinsic stellar variability, early studies suggest that there is sufficient information to measure velocities much more precisely than using traditional methods, with reductions in the RV rms due to activity of up to 50% demonstrated to date (e.g., Davis et al. 2017, Lanza et al. 2018) . Recognizing the need to measure the masses of Earth-like planets orbiting Sun-like stars, the NAS Exoplanet Science Strategy report recommended, *"NASA and NSF should establish a strategic initiative in extremely precise radial velocities (EPRVs) to develop methods and facilities for measuring the masses of temperate terrestrial planets orbiting Sun-like stars."* It is therefore not unreasonable to theorize that potentially habitable planets orbiting the nearby Sun-like stars will be detected by RV surveys in advance of future direct imaging missions.

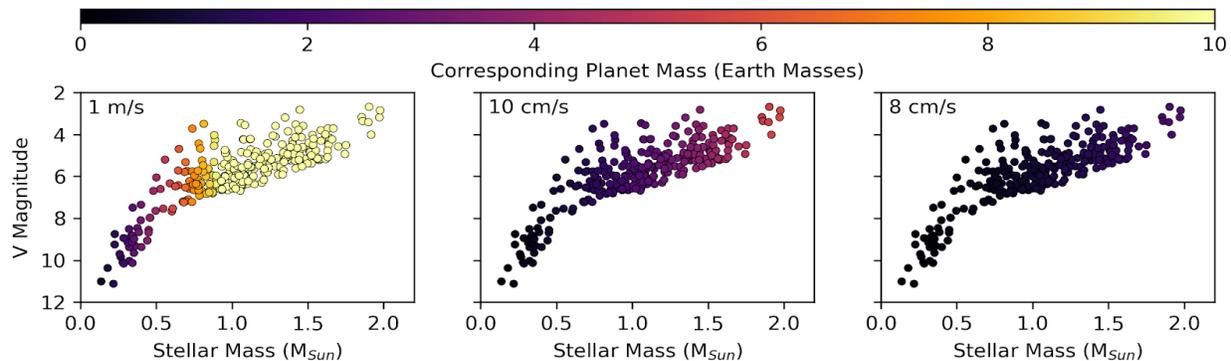

*Figure 2: Anticipated RV semi-amplitude generated by the presence of an Earth-mass planet receiving the same insolation flux as the Earth versus V-band magnitude for stars with HZs accessible to direct imaging with the ECLIPS coronagraphic instrument on LUVOIR-A. The points are colored to indicate the mass of the host star (left) and the orbital period of the planet (right). Figure based on simulations completed by Chris Stark (STScI).*

For the most massive stars, the anticipated RV signals are small (3.5 - 5 cm/s) and occur over long timescales (2 - 4 years), implying that Earth-like planets would be challenging to detect even with extremely precise spectrographs on the ELTs. However, RV observations will still be useful for identifying massive planets that could be confused for Earth-mass planets. Although multi-band photometry or spectroscopy would in principle distinguish these false positives from true candidate Earth-like planets, there may be difficult cases, such as a Neptune-mass planet in the HZ seen at crescent phase, which might have similar photometry or spectra to a terrestrial planet. In addition, multi-band photometry will not resolve the degeneracy in all cases and spectroscopy is time-intensive. Obtaining RV observations in advance would improve the efficiency of a future imaging survey by reducing the amount of time spent distinguishing terrestrial planets from more massive interlopers with similar orbital parameters.

As shown in *Figure 3*, a survey capable of detecting an RV signal of 1 m/s on the timescale of HZ orbital periods could detect Neptune-mass planets, 10 $M_{Earth}$ planets, and 3 $M_{Earth}$ planets in the HZs of F stars, G stars, and M dwarfs, respectively. Improving the precision to 20 cm/s would permit the detection of 4 $M_{Earth}$, 2 $M_{Earth}$, and 0.5 $M_{Earth}$ planets. Further enhancing the performance to 8 cm/s would reveal 1.4 $M_{Earth}$, 0.9 $M_{Earth}$, and 0.2 $M_{Earth}$ planets.

**Improving the Efficiency of a Future Direct Imaging Survey**

In the absence of advanced knowledge about which nearby stars host planets, a future direct imaging mission would need to conduct an initial search to find planets before obtaining spectra to determine the properties. In the opposite extreme, if the planet census was already complete, then the detection stage could be bypassed entirely and the full mission could be devoted to planet characterization after taking initial images to confirm the positions and brightness of the previously-detected planets. For LUVOIR-A/ECLIPS, simulations by C. Stark indicate that the search phase would comprise roughly 71% (1.4 years) of a 2-year campaign devoted to detecting and characterizing Earth-like planets (Roberge & Moustakas 2018). Similarly, C. Stark predicts that a mission using an observatory similar to the HabEx mission concept study would spend roughly 1.2 years out of a 2.5-year campaign searching for planets prior to characterization (Gaudi et al. 2018).

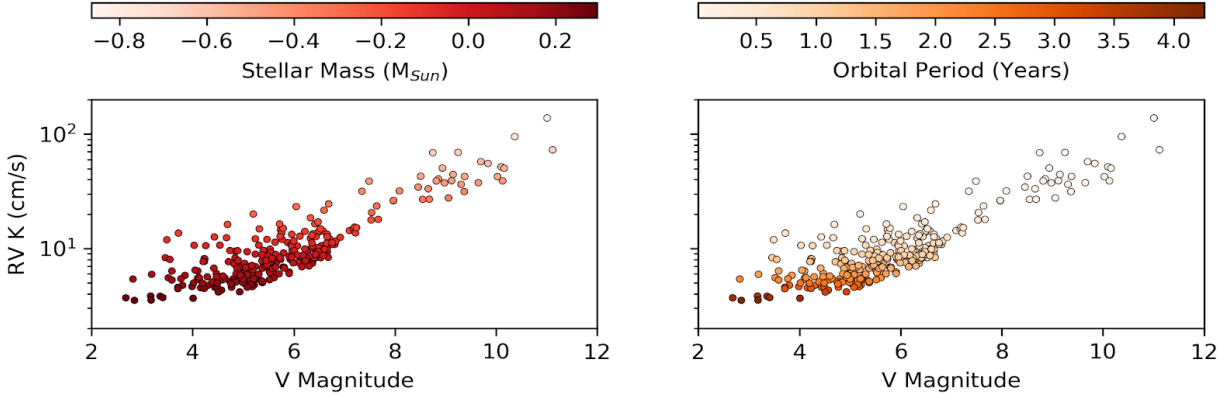

*Figure 3: Minimum detectable mass of an HZ planet as a function of stellar mass and V-band magnitude. The color-coding indicates the planet mass corresponding to an RV semi-amplitude of 1 m/s (left), 10 cm/s (middle) or 8 cm/s (right). In order to enhance the detail in the rightmost panels, all planets more massive than 10 Earth masses are shown in yellow.*

A dedicated RV campaign could find a subset of the best imaging targets in advance. *Figure 4* displays five possible time breakdowns corresponding to varying levels of knowledge about the population of nearby Earth-like planets. The most pessimistic and optimistic scenarios assume complete ignorance or complete knowledge about which stars host Earth-like planets. The three intermediate scenarios assume knowledge of all planets that induce RV signals larger than 20 cm/s, 10 cm/s, or 8 cm/s. Compared to the most pessimistic case of no prior knowledge, the time required for initial imaging is reduced by 9%, 25%, and 44%, respectively. The revised time estimates include a single image to verify the position and brightness of the planet before acquiring spectra for atmospheric characterization. Reducing the search phase will increase the time available for characterization at the cost of decreasing the number of serendipitous detections of additional planets and disks. The time saved might therefore be used to conduct a broader survey of system demographics at lower contrast (e.g., Astro2020 WP by M. Marley).

Knowledge of planet masses and orbits will enable target prioritization so that the targets most likely to be rocky and amenable to life could be observed first and at key orbital phases. A planet detected by an uninformed imaging search requires detections at least three epochs to accurately determine its orbital parameters, whereas the combination of RV data and 1-2 well-timed images can fully constrain planetary orbits and confirm their position within the HZ.

Mass constraints would also be useful for identifying the best targets for observations with the ELTs. As described in the Astro2020 WP by T. Currie and J. Wang, the ELTs will be capable of obtaining thermal IR images of Earth-like planets in the HZs of AFGK stars and shorter wavelength image of Earth-like planets orbiting M dwarfs. These observations will complement the visible and NIR spectra acquired by future space-based direct imaging missions.

In particular, ELT observations at 10 microns could constrain planet radii as long as the temperature can be estimated. Combining radius estimates from thermal IR imaging with mass estimates from RV observations would yield planet densities, thereby constraining bulk compositions and revealing which planets are least likely to be low-density planets.

**Understanding Atmospheric Spectra and Potential Biosignatures**

Planet mass measurements are essential for accurately interpreting planetary spectra because there is a degeneracy between surface gravity and the abundance of atmospheric constituents (Batalha et al. 2017, Nayak et al. 2017, Astro2020 WP by Ciardi). In an ideal case, the planet mass would be constrained accurately and precisely by indirect methods, but even upper limits on planet mass would restrict the allowed parameter space and enable better constraints on atmospheric composition. For transiting planets, radii can be easily calculated from the observed transit depth and the stellar radii. However, most Earth-like planets observed by direct imaging missions will not transit their host stars because the geometric likelihood of transit decreases with increasing orbital period. For non-transiting planets, radii can be estimated from the observed brightness and informed estimates of planet temperature and albedo.

**Conclusions and Recommendations**

Future space-based missions will be capable of searching for biosignatures in the atmospheres of potentially habitable planets, but those planets must be identified before they can be characterized. Although future missions could conduct uninformed surveys for Earth-like planets, conducting a precursor RV survey of the best target stars would maximize the scientific return of a future direct imaging mission by finding candidate Earth-like planets, revealing more massive planets that could be initially mistaken for Earth-like planets, and providing the essential surface gravity constraints required to interpret planetary spectra. Constraints on planet masses will be required regardless of whether the planets are detected directly or indirectly. Given the long orbital timescales of Earth-like planets orbiting Sun-like stars, several years of data will be required to determine orbits. Accordingly, **we recommend an intensive effort to conduct the deepest RV survey possible of the stars whose habitable zones are the most amenable to direct imaging.** Successfully accomplishing this survey will require highly-stabilized spectrographs to obtain an exquisite data set of frequent and numerous observations at high spectral resolution and high signal-to-noise as well as advanced statistical methods to analyze the resulting set of observations. This effort is beyond the scope of any single observatory or RV group, and would require significant investment in both facilities and human resources.

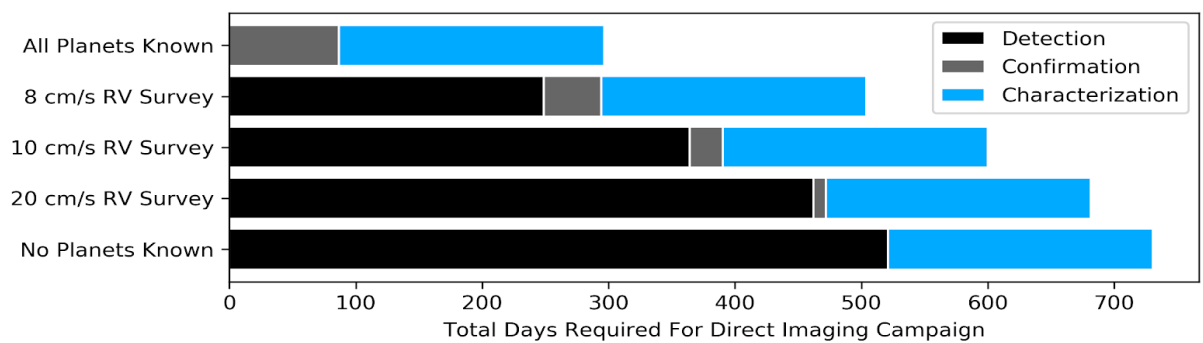

*Figure 4: The time saved by detecting planets in advance. The bars compare the time spent conducting an initial imaging search for planets (black), obtaining a single image to verify planet positions (grey), or acquiring spectra to characterize planets (blue).*